# Drift-Aware Explainable Prioritization for Windows PE Malware Forensic Triage with Semantic Consistency Checks

Mohammed A. Alharbi
Faculty of Computer and Information Systems
Islamic University of Madinah
471020826@stu.iu.edu.sa

Toqeer Ali Syed
Faculty of Computer and Information Systems
Islamic University of Madinah
toqeer@iu.edu.sa

**Abstract**

Forensic investigators must decide which suspicious files to examine first. Machine-learning (ML) and deep-learning (DL) malware detectors usually return a bare score that neither explains why a file deserves attention nor signals when changing data make the model unreliable. We present a framework for Windows Portable Executable (PE) malware forensic triage that works only on pre-extracted static features and never executes malware. A calibrated maliciousness score sets each file's triage priority, and a per-sample feature attribution explains the prediction. A per-sample consistency score measures whether the top attributed features agree with documented behaviors from the Malware Behavior Catalog (MBC) and MITRE Adversarial Tactics, Techniques & Common Knowledge (ATT&CK), where the available features support such interpretation. Conformal evaluation and statistical distribution tests flag drift. Each file receives one investigator-oriented triage output. The evaluation protocol, specified before testing, uses a time-ordered split of EMBER2024 as the primary dataset and BODMAS for family-disjoint checks, and compares every component with simple baselines on predictive performance, prioritization effectiveness, explanation quality and robustness over time.

## 1 Introduction

Digital forensics and incident response teams receive more suspicious files than they can examine in depth. Triage deliberately trades analytical depth for speed: the field triage model of Rogers et al. identifies and interprets digital evidence within a short time frame without first acquiring full images [1], and in malware analysis triage accepts lower accuracy in exchange for faster prioritization [2]. In this paper, *malware triage* means the automated ranking of incoming Windows Portable Executable (PE) files so that investigators examine the highest-risk files first.

Machine learning (ML) and deep learning (DL) detectors are natural candidates for this ranking. Learning-based malware detection has been surveyed extensively [2, 3]; gradient-boosted trees on parsed static features detect PE malware [4]. A detector's score alone, however, does not meet an investigator's needs in three respects.

**Explanation.** An investigator must be able to justify why a file was prioritized [5, 6]. Hall et al. argue that artificial intelligence (AI)-assisted decisions in fields such as law enforcement need to be justifiable and explainable to a human [7]. Post-hoc attribution methods such as Local Interpretable Model-agnostic Explanations (LIME) [8] and SHapley Additive exPlanations (SHAP) [9] are applied to malware detectors [10, p. 61634]; Integrated Gradients (IG) [11] is a related attribution method. Their quality has to be measured on explicit criteria [12], and explanations in adversarial environments can be manipulated [13].

**Semantics.** Attributions name low-level features, such as header fields, section properties or imported functions [14], not behaviors. Taxonomies such as Adversarial Tactics, Techniques & Common Knowledge (ATT&CK) [15] and Malware Behavior Catalog (MBC) [16] describe behaviors in terms that investigators already use. Among the works reviewed here [10, 17, 18, 19, 20, 21, 22, 23, 24], we did not identify a per-sample measure comparing a PE maliciousness detector's static feature attributions with that file's MBC/ATT&CK reference technique set. Guldemir et al.'s capa and ATT&CK analysis is independent of their feature-attribution pipeline [23, p. 2]. Švec et al. map imported functions to Malware Attribute Enumeration and Characterization (MAEC) malware actions [25] in an ontology over static PE features [17]. DeepReflect evaluates the localization of malicious functions within PE binaries against analysts' MITRE ATT&CK labels [18, p. 3477]; this concerns disassembled functions rather than a detector's static feature attributions. Dream explains Android drift with its own behavior concepts and envisages ATT&CK-based concepts only as an extension [19, p. 989].

**Drift.** Malware and benign software change over time, and concept drift degrades detectors [26, 27]. Chen et al. report that an Android malware classifier trained on one year of data saw its balanced F-score (F1) drop from 0.99 to 0.76 after six months [28], and evaluations that ignore time inflate results [29, 30]. Drift detection [31, 32, 19] and drift explanation [32, 19] exist, but the rankings these systems produce serve labeling budgets or drift inspection rather than providing a calibrated triage priority over all incoming files [32, 28, 19].

**Approach.** We design an integrated triage framework for static Windows PE features that combines a calibrated triage priority, per-sample attributions, per-sample agreement with MBC/ATT&CK reference techniques, and drift flags; unlike Guldemir et al., who analyze drift explanations and behavioral profiles separately [23, pp. 2, 7–9], the semantic check operates on each file. A calibrated detector provides the triage priority; a post-hoc explainer attributes each prediction to features; a documented feature-to-behavior mapping turns the top attributions into MBC and ATT&CK identifiers whose agreement with reference tags yields a per-sample consistency score; and conformal evaluation and distribution tests flag cases and periods that may require further investigation or model reassessment. All four signals are combined in one investigator-oriented triage output per file.

**Contributions.**

- A triage framework on pre-extracted static Windows PE features, without executing malware, centered on per-sample agreement between maliciousness attributions and MBC/ATT&CK reference technique sets, alongside calibrated triage priority and drift awareness (Section 4); this agreement comparison is distinct from Guldemir et al.'s independent analysis of drift partitions [23, pp. 2, 8–9].
- An agreement measure between per-sample feature attributions and documented MBC/ATT&CK behaviors, with a per-sample consistency score, coverage reporting and a permuted-mapping null baseline (Sections 4 and 5).
- A time-ordered evaluation protocol on EMBER2024 [33], a new dataset created in collaboration with the authors of EMBER2017 and EMBER2018. EMBER2017 introduced the Endgame Malware BEnchmark for Research (EMBER) [4]. BODMAS (Blue Hexagon Open Dataset for Malware AnalysiS), an open dataset of timestamped Windows PE malware [34], is used as a secondary dataset. The protocol compares each component with simple baselines under leakage controls and with parameters fixed before testing (Section 5).

# 2 Background and Related Work

## 2.1 Static analysis of Windows PE malware

Static analysis inspects a file's content without running it, whereas dynamic analysis runs the file to observe its behavior [2]. Static analysis has known limits [35]: Moser et al. show that obfuscation based on opaque constants defeated virus scanners and a static detector, and conclude that static analysis alone may not be enough [36]. The EMBER dataset provides features extracted from 1.1 million PE files, and its authors report that a LightGBM baseline with default settings slightly outperformed a MalConv model [37] that they trained on the same files [4]. EMBER2024 extends this line with a benchmark for holistic evaluation of malware classifiers [33]; its authors suggest that the EMBER feature version 3 features are "resistant to concept drift due to small temporal changes" [33, p. 5521]. BODMAS is a family-labeled dataset of PE malware for temporal analysis [34].

## 2.2 Explainable AI for malware analysis

SHAP is a unified framework of additive feature attributions [9], LIME learns an interpretable model locally around a prediction [8], and IG attributes deep-network outputs under the Sensitivity and Implementation Invariance axioms [11]. Manthena et al. survey these and other methods for malware analysis [10, pp. 61618–61621]. Warnecke et al. propose general criteria (descriptive accuracy and sparsity) and security-specific criteria (completeness, stability, robustness and efficiency) for comparing explanation methods [12]. Noppel and Wressnegger systematize how explanations can be forged through input and model manipulations [13]. Downing et al. localize malicious functionality within PE binaries with an autoencoder [18]. Švec et al. build a semantic representation of static PE features for explainable detection [17]. Partition-based Tree Explanation (PaTEx) clusters malware samples to partition the input space of a neural malware classifier into decision-aware regions and learns a local surrogate decision tree for each region [20, pp. 5882–5884]; on three datasets (an EMBER subset of static PE features, a dynamic Windows dataset and an Android dataset), its authors report higher fidelity to the neural model than SHAP and LIME [20, pp. 5884, 5886].

On CICMalMem22 memory-forensic features and a refined CICIDS17 network-flow release, AL-Essa et al. report changes in SHAP and IG attributions even when the deep neural network (DNN)'s predicted label is preserved [24, pp. 10–12, 15]. Their comparison concerns clean and perturbed versions of the same inputs, rather than chronological drift in static PE features [24, pp. 9–12, 23].

## 2.3 Concept drift in malware detection

Gama et al. define concept drift as a change over time in the relation between input data and the target variable [26]. TESSERACT shows that spatial and temporal bias inflate the results of Android malware classifiers [29], and Arp et al. describe temporal snooping among the common pitfalls of ML in security [30]. Conformal evaluation underpins classification with rejection: Transcend [38] and Transcendent [31] withhold low-confidence predictions under drift. Contrastive Autoencoder for Drifting detection and Explanation (CADE) detects drifting samples and explains why they are new [32]; Dream uses behavior concepts to detect, explain and adapt to drift for Android classifiers [19]; Li et al. adapt a graph neural network detector to drifted Windows malware from few samples [39]; and Chen et al. study continuous learning for Android malware [28]. Two rule-based approaches address drift interpretably: Kalný et al. train decision-tree classifiers across temporal windows of EMBER2024 for six malware families and quantify drift by comparing the extracted rulesets, with Repeated Incremental Pruning to Produce Error Reduction (RIPPER) and Transcendent as baselines [21, pp. 922–923]; and Jurečková and Jureček compare RIPPER rule sets learned on original samples of six malware families and on synthetically evolved variants of them, to detect drift and identify the features whose values changed [22, pp. 366–370]. Continual-learning methods reduce the cost of keeping detectors current: Malware Analysis with Distribution-Aware Replay (MADAR) replays a mix of representative samples and outliers selected with isolation forests and is evaluated on EMBER Windows data and an Android benchmark [40, Sec. 1].

Guldemir et al. explain a triplet-autoencoder and Density-Based Spatial Clustering of Applications with Noise (DBSCAN) drift score with IG on a time-ordered BenchMFC subset of known Windows PE malware families represented by static EMBER features [23, pp. 5–9]. Their Drift Correction Rate (DCR) assesses feature-level consistency with the detector's drift decision, whereas Jaccard similarity and Jensen–Shannon divergence compare capa capability and ATT&CK profiles aggregated over detector-defined partitions [23, pp. 7–9, 14–15]. Behavioral separation varies across families, so these partition comparisons provide complementary evidence rather than universal validation [23, pp. 14–15].

### 2.4 Behavior taxonomies

MITRE ATT&CK is a knowledge base of adversary tactics and techniques built from real-world observations [15]. MBC catalogs malware objectives and behaviors to support labeling and reporting, with links to related ATT&CK techniques where appropriate [16].

### 2.5 Positioning

Related approaches include Guldemir et al. for Windows PE drift-score explanation and independent ATT&CK profile analysis [23, pp. 2, 5–9], DeepReflect for function localization on PE [18], Dream for behavior-level drift explanation on Android [19], and CADE for drift detection and feature-level explanation [32]. Each addresses only part of the triage problem. Our framework differs in producing, for static PE features, a single triage output that combines a calibrated triage priority, per-sample attributions, a per-sample agreement measure with MBC/ATT&CK, and case- and period-level drift flags; Guldemir et al. explain drift at the feature level and analyze ATT&CK profiles at the partition level, separately from their attributions [23, pp. 2, 7–9]. We make no claim beyond the works reviewed here.

## 3 Threat Model and Problem Formulation

### 3.1 Setting and threat model

An investigator receives a stream of Windows PE files in weekly batches and can examine only a limited number in depth. The framework sees each file only as its pre-extracted static feature vector and never executes it. Malware authors adapt their techniques over time [31, p. 805]. Concept drift is a change in the input–target relation [26]; the framework must therefore signal when its scores may no longer be reliable. An adaptive adversary may also try to manipulate detectors or explanations [13, 41, 42]. We treat such manipulation as a stated limitation: we neither evaluate robustness against it nor describe any procedure for it. Antivirus-derived labels can be imperfect on Windows PE data [43, p. 3996]. Labels and reference behavior tags come from the datasets and are not ground truth; agreement with them indicates explanation plausibility, not correctness. Explanations are presented as a guide to inspection [44, p. 2078], not as evidence in their own right.

## 3.2 Problem

Given a time-ordered stream of static PE feature vectors, produce for each file a calibrated priority, an explanation of the features contributing to it, an assessment of whether that explanation is consistent with documented malware behaviors where the features support such interpretation, and a drift flag for files or periods that may require further investigation or model reassessment.

## 3.3 Research questions

The study is framed by one design-science question and five sub-questions, quoted verbatim.

**Main RQ.** *How can an AI-assisted malware forensic triage framework that integrates machine-learning and deep-learning (ML/DL) prioritization, explainability, semantic assessment of whether explanations are consistent with documented malware behaviors (MBC/ATT&CK, scored per sample and per technique where the available features support such interpretation), and drift awareness be designed, implemented, and evaluated so that it provides an investigator-oriented triage output from pre-extracted static features of Windows Portable Executable files without executing malware, and how does it compare with selected baseline approaches in predictive performance, prioritization effectiveness, explanation quality, and robustness over time?*

**SRQ1.** *Under a time-ordered evaluation on EMBER2024 Windows PE static features, to what extent does a calibrated ML/DL maliciousness score improve predictive performance and prioritization effectiveness compared with selected baseline approaches?*

H0a: "$\Delta$area under the precision–recall curve (PR-AUC) (proposed $-$ strongest baseline) $= 0$ on the temporal test set. H1a: $\Delta$PR-AUC $\neq 0$ (expected $> 0$)." H0b: "At the threshold fixed on validation (true positive rate (TPR)=95%), the two models have equal error rates, i.e. their discordant decisions are symmetric. H1b: they are not."

**SRQ2.** *Which post-hoc explanation method gives the most faithful and stable feature attributions for the SRQ1 detector, compared with a simple baseline explanation, and does explanation quality degrade on later (drifted) test periods?*

H0a: "No difference in deletion-based fidelity between methods (e.g. TreeExplainer (Tree SHAP) vs. LIME vs. Integrated Gradients for the DL model)." H0b: "Fidelity and stability on the final test period equal those on the first test period."

**SRQ3.** *To what extent are the top attributed features consistent with documented malware behaviors (MBC/ATT&CK), where the available features support such interpretation, and does a per-sample consistency score separate correct from incorrect high-priority predictions?*

Part (a) is descriptive. Part (b): "H0: The per-sample consistency score does not discriminate correct from incorrect predictions (area under the receiver operating characteristic (ROC) curve (AUROC) $= 0.5$). H1: AUROC $> 0.5$."

**SRQ4.** *How early and how reliably do conformal evaluation and statistical distribution tests flag cases or periods that may require further investigation or model reassessment on time-ordered EMBER2024 data, compared with a performance-based retraining trigger?*

H0: "Detection delay (weeks from onset of significant performance drop to alarm) is not shorter for detector D than for the performance-based trigger at a matched false-alarm rate. H1: D is shorter."

**SRQ5.** *What should an investigator-oriented triage output contain so that priority, explanation, semantic-assessment status and drift flag are interpretable and actionable, and how well does the implemented prototype provide it?* This question follows design science and is not hypothesis-testable as a whole.

The study objectives are, verbatim: O1 "To design an AI-assisted malware forensic triage frame-work integrating prioritization, explainability, semantic assessment, and drift awareness.”; O2 “To implement a prototype that provides investigator-oriented triage results.”; O3 “To evaluate the framework’s performance, explanation quality, and robustness over time.”; O4 “To compare the framework with selected baseline approaches for malware forensic triage.”

# 4 Framework

The framework takes the pre-extracted static feature vector of a Windows PE file and returns one triage record per file. It has four components, each with its own evaluation (Table 1); Figure 1 gives an overview.

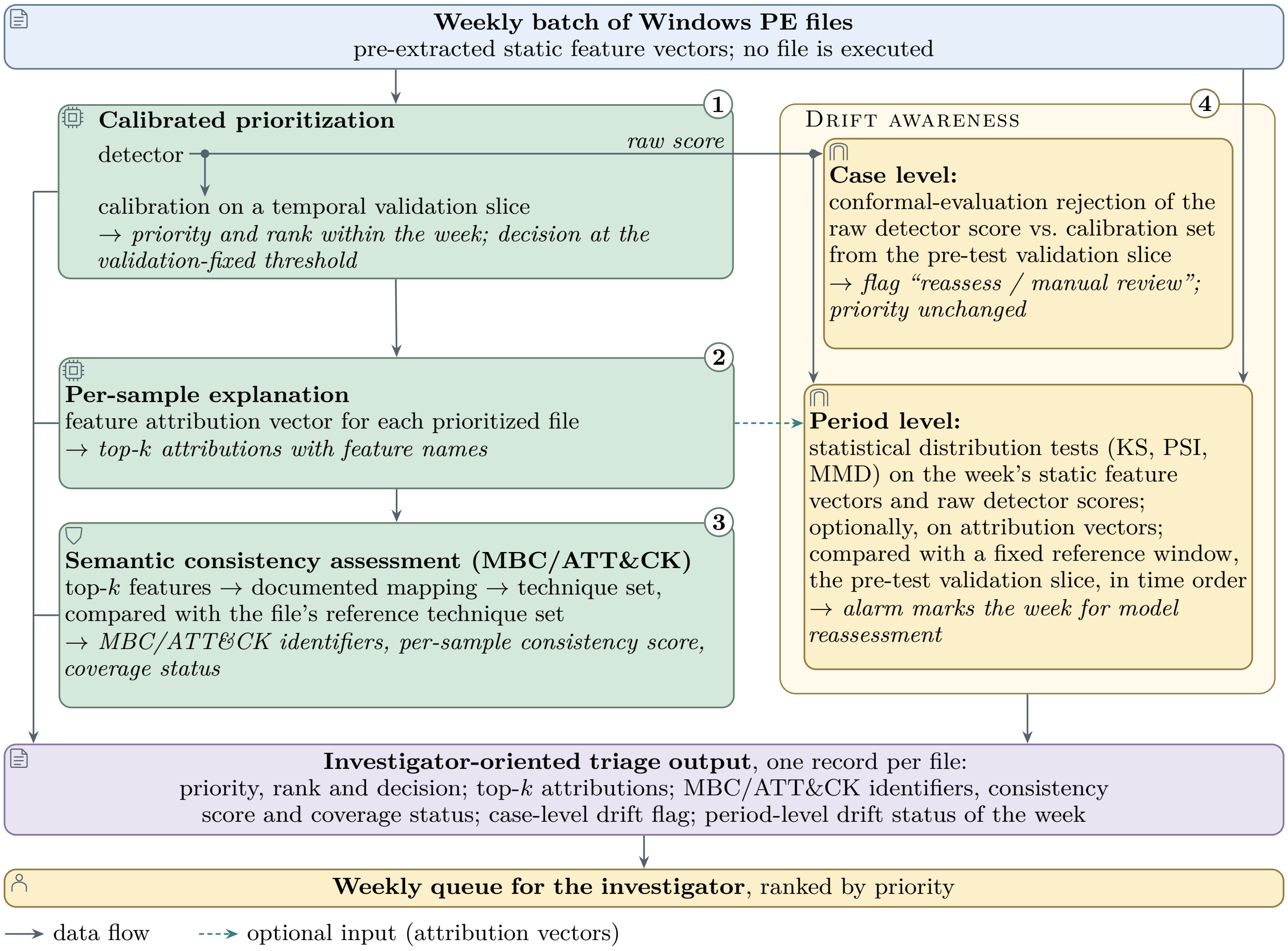

Figure 1: Conceptual overview of the framework. Each file in a weekly batch passes through calibrated prioritization (1), per-sample explanation (2) and semantic consistency assessment (3); drift awareness (4) flags files by conformal-evaluation rejection and weeks by distribution tests against the pre-test validation slice. The output is one triage record per file in a priority-ranked weekly queue. The diagram shows no data or results.

**Calibrated prioritization.** A detector outputs a maliciousness score. The score is calibrated with Platt scaling or isotonic regression [45, 46], fitted only on a temporal validation slice that precedes the test period, which avoids data snooping (pitfall P3 in [30, p. 3973]). The calibrated probability is the triage priority, and files are ranked within each weekly batch. A decision threshold is fixed on validation at a TPR of 95% and is not re-tuned on test data.

**Per-sample explanation.** Each prioritized file receives a feature attribution vector. The candidate methods are TreeExplainer for the tree model [47], LIME [8], and IG for the DL model [11]. The evaluation criteria are adapted from Warnecke et al. [12, pp. 162–164].

**Semantic consistency assessment (MBC/ATT&CK).** Our mapping is designed to link interpretable static features, for example import and application programming interface (API)-name features and section features, to MBC behaviors and ATT&CK technique identifiers. MBC supplies a malware-analysis behavior vocabulary and documents associations of capa rules with MBC behaviors and/or ATT&CK techniques [16]. ATT&CK supplies the technique taxonomy [15]. Two coders build the mapping independently, and Cohen's $\kappa$ is reported [48]. The mapping covers only features that admit a behavioral reading, and its coverage is reported; this implements the restriction "where the available features support such interpretation". capa rule files [49] serve only as documentation for the mapping; capa is not executed. For each file, the top-$k$ attributed features are mapped to a technique set. The agreement measure compares this set with the file's reference technique set through per-sample precision and recall. For file $i$, let $\mathcal{T}_i^{\text{map}}$ be the mapped technique set and $\mathcal{T}_i^{\text{ref}}$ the reference technique set. The *per-sample consistency score* is the harmonic mean of per-sample precision and recall, which equals the Dice coefficient of the two sets:

$$C_i = \frac{2\,|\mathcal{T}_i^{\text{map}} \cap \mathcal{T}_i^{\text{ref}}|}{|\mathcal{T}_i^{\text{map}}| + |\mathcal{T}_i^{\text{ref}}|}. \tag{1}$$

Files with an empty mapped set are reported as not interpretable from available features; files with an empty reference set are excluded from the score, and their count is reported with the results. Agreement is summarized per technique as a macro average over techniques with at least 30 reference-tagged samples. Figure 2 summarizes the semantic comparison and the handling of features without a behavioral reading.

**Drift awareness.** At case level, conformal-evaluation rejection in the Transcend/Transcendent family [38, 31] ([31, Sec. III, p. 807]) flags rejected files for reassessment or manual review [50]. Its input is the detector's output for the file: a non-conformity score derived from the raw (uncalibrated) maliciousness score for the predicted class, compared with the non-conformity scores of a calibration set drawn from the pre-test validation slice; feature vectors enter only through the detector, and attribution vectors are not an input. This flag is the framework's only abstention mechanism and does not alter priority. At period level, statistical distribution tests run on features or scores, and optionally on attribution vectors: the Kolmogorov–Smirnov (KS) test [51], the population stability index (PSI) [52] and maximum mean discrepancy (MMD) [53]. Each test week's batch (its static feature vectors, its raw detector scores and, optionally, its attribution vectors) is compared with the same quantities on a fixed reference window, the pre-test validation slice, in time order, and an alarm marks the week for model reassessment. The calibration set, the reference window and the alarm thresholds are design choices, fixed before testing and not tuned.

**Triage output.** The investigator-oriented triage output contains one record per file with: the priority and rank; the decision at the validation-fixed threshold; the top-$k$ attributions with feature names; the mapped MBC/ATT&CK identifiers with the consistency score and coverage status ("not interpretable from available features" where applicable); the case-level drift flag; and the period-level drift status of the file's week.

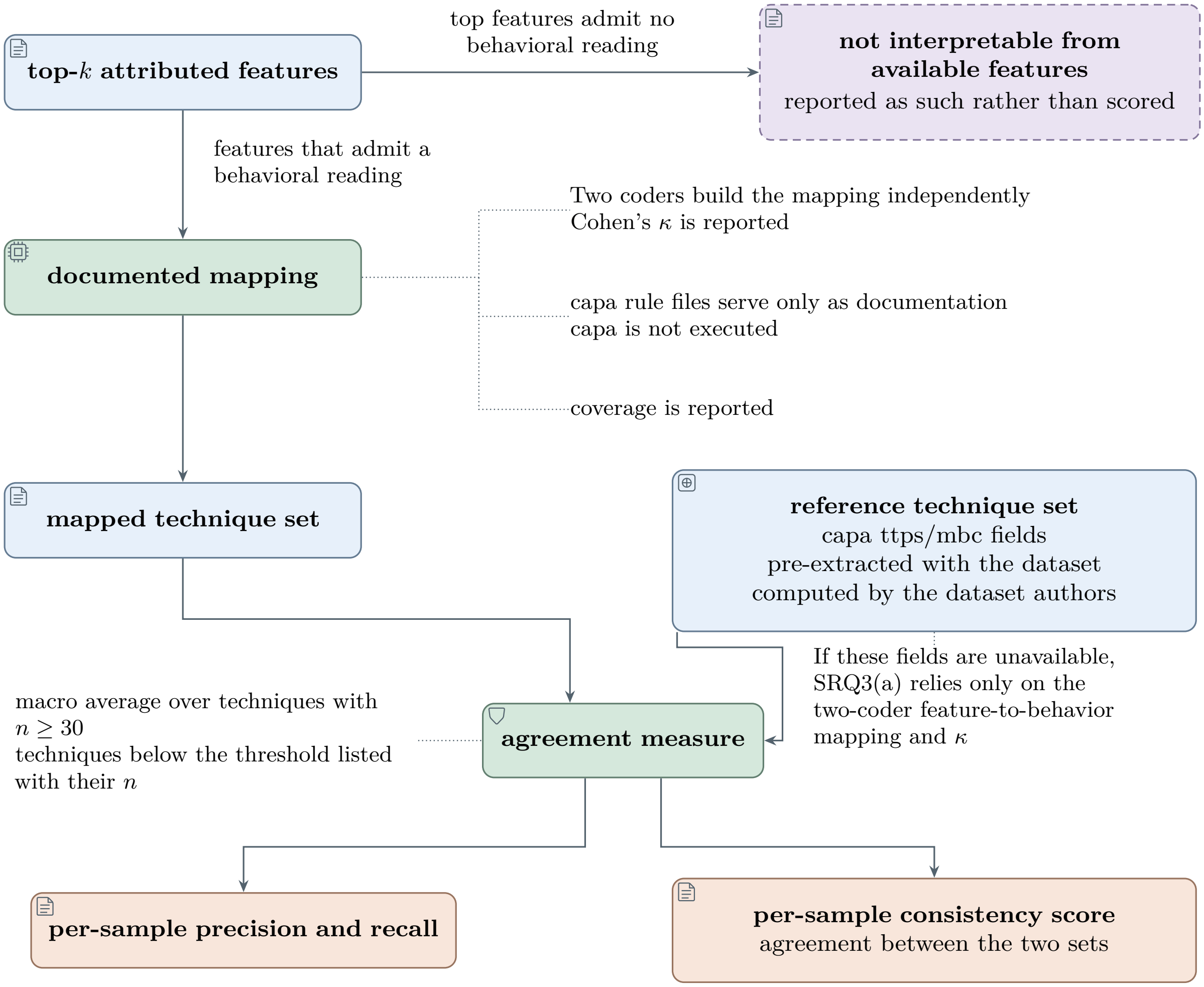

Figure 2: Semantic consistency assessment for the PE subset. Top-$k$ attributed features that admit a behavioral reading pass through a documented mapping to a technique set. The agreement measure compares this set with the file's pre-extracted capa ttps/mbc reference technique set through per-sample precision and recall and a per-sample consistency score. Files whose top features admit no behavioral reading are reported as not interpretable from available features rather than scored. ClarAVy agreement is reported separately per tag, never in the per-technique average.

The metrics in Table 1 include area under the curve (AUC) measures: the PR-AUC and the AUROC.

Table 1: Component-to-evaluation map.

| Component | SRQ | Evaluation dimension | Primary metric | Baselines |
|---|---|---|---|---|
| Prioritization | SRQ1 | predictive performance; prioritization effectiveness | PR-AUC; workload | B0–B3 |
| Explanation | SRQ2 | explanation quality | deletion fidelity | E0, E1 |
| Semantic | SRQ3 | explanation quality | per-sample precision/recall; AUROC | S0 |
| Drift | SRQ4 | robustness over time | detection delay | D0–D2 |
| Output | SRQ5 | (O2) | requirement coverage | – |

# 5 Experimental Setup

## 5.1 Datasets

All analyses use pre-extracted static features and released labels. No binaries, raw bytes or disassembly are obtained, and no file is executed.

**EMBER2024 (primary) [33, 54].** The dataset covers Windows PE files with static features from September 2023 to December 2024. We use its PE subset (Win32, Win64 and .NET) with the split as released: the first 52 weeks for training and the final 12 weeks for testing. The PE subset comprises 2,340,000 training and 540,000 test files, with 45,000 PE files per week. Family labels and behavior tags come from ClarAVy [55], and the capa `ttps/mbc` fields used as reference technique sets ship pre-extracted with the dataset and were computed by the dataset authors [54]. If these fields are unavailable for the PE files in the release used, SRQ3(a) relies only on the two-coder feature-to-behavior mapping and $\kappa$. The challenge set is not used.

**BODMAS (secondary) [34].** BODMAS provides 57,293 malware files collected from August 2019 to September 2020 and 77,142 benign files collected from January 2007 to September 2020, with 581 analyst-curated families [34, pp. 78–79]; its features follow the format of EMBER and Sophos/ReversingLabs-20 Million (SOREL-20M) [34, p. 80]. Only the pre-extracted feature vectors are used; the malware binaries distributed with BODMAS are not obtained. It is used for family-disjoint checks for SRQ1, reported separately, and for a secondary SRQ4 analysis specified before testing, with its own onset criterion, run only if no drift onset occurs in the EMBER2024 test weeks. Because BODMAS benign files span 2007–2020, the secondary SRQ4 analysis uses only benign files dated within the malware window, and their count is reported. Because the feature versions of the two datasets differ, they are never pooled and no model is transferred between them.

**Terms of use.** EMBER2024's code (GitHub FutureComputing4AI/EMBER2024, commit 0ef753e8) and its released features and labels (Hugging Face joyce8/EMBER2024) are both distributed under the Apache License 2.0, and the dataset is cited as requested by its authors. BODMAS feature vectors and metadata are publicly downloadable from the authors' site (code under the 2-Clause Berkeley Software Distribution (BSD) license, BSD-2-Clause), malware binaries are shared only on request under a no-onward-sharing condition, benign binaries are not released, and the dataset is cited as requested.

## 5.2 Temporal protocol and leakage controls

- *Time ordering.* Every training object strictly precedes every test object [29, p. 734]. The time key is the released VirusTotal first-submission week.
- *Validation.* The final $w_{\text{val}}$ weeks of the training period ($w_{\text{val}}$ is fixed before testing and reported with the results) are used for early stopping, calibration, threshold fixing and conformal calibration. Test weeks are never used for fitting or tuning (pitfalls P3 and P5 in [30, pp. 3973–3974]).
- *Window consistency.* Within each test week, benign and malicious files come from the same time window, which the released weekly sampling guarantees [33, p. 5517].
- *Malware-to-goodware ratio.* The released test set is balanced at 50% malware; this is relevant to the base-rate fallacy (pitfall P8 in [30, p. 3975]). Primary analyses use the test set as released. A sensitivity analysis specified before testing repeats the SRQ1 metrics after subsampling malware in each test week to a realistic malware-to-benign ratio. The ratio is treated as a sensitivity range $\pi$, fixed before testing and reported with the results, that includes about 1:55, the ratio of the evaluation set of Huynh et al., which they describe as close to the malware prevalence reported in enterprise endpoint telemetry and as a plausibility anchor rather than a universal value [56, p. 4]. Benign files are kept in full, and the subsampling is repeated $r$ times with fixed seeds; $r$ is fixed before testing and reported with the results.
- *Random-split contrast.* To show how much temporal snooping inflates results, one stratified random split of the same files is evaluated with the identical pipeline [30, 29].
- *Fitting scope.* Feature scaling, calibrators and mapping thresholds are fitted on training or validation data only.
- *Deduplication.* Exact hash duplicates across the training and test sets are removed. Near-duplicates across weeks are not removed.
- *Seeds.* At least five seeds are used per model, and the per-seed spread is reported alongside seed-averaged estimates.
- *Parameters fixed before testing.* The weekly capacity $K$, the attribution top-$k$, the per-technique threshold ($n \geq 30$), the onset criterion, $\pi$, $w_{\text{val}}$, the D1 schedule and the matched false-alarm rate targets are fixed before the test set is first scored. Design choices not taken from the cited sources (at least five seeds, the per-technique threshold $n \geq 30$, D1 retraining every 4 test weeks and the Likert scale midpoint 3) are fixed before testing and are not tuned.

Figure 3 summarizes the training, validation and test periods and the uses of the pre-test validation slice.

## 5.3 Models and baselines

Table 2 lists the proposed models and baselines, each with a unique identifier (ID). Simple, well-understood baselines address pitfall P6 in [30, pp. 3974–3975]. DeepReflect [18] and FINER [57] are reference points outside our pre-extracted EMBER-style feature setting. CADE [32] is an optional reference baseline in our protocol; we do not re-evaluate it on EMBER2024 features.

## 5.4 Metrics

- *Predictive performance (SRQ1).* PR-AUC; F1 at the validation-fixed threshold; false positive rate (FPR) at TPR = 95% (FPR@TPR95), with the realized TPR per week; and expected

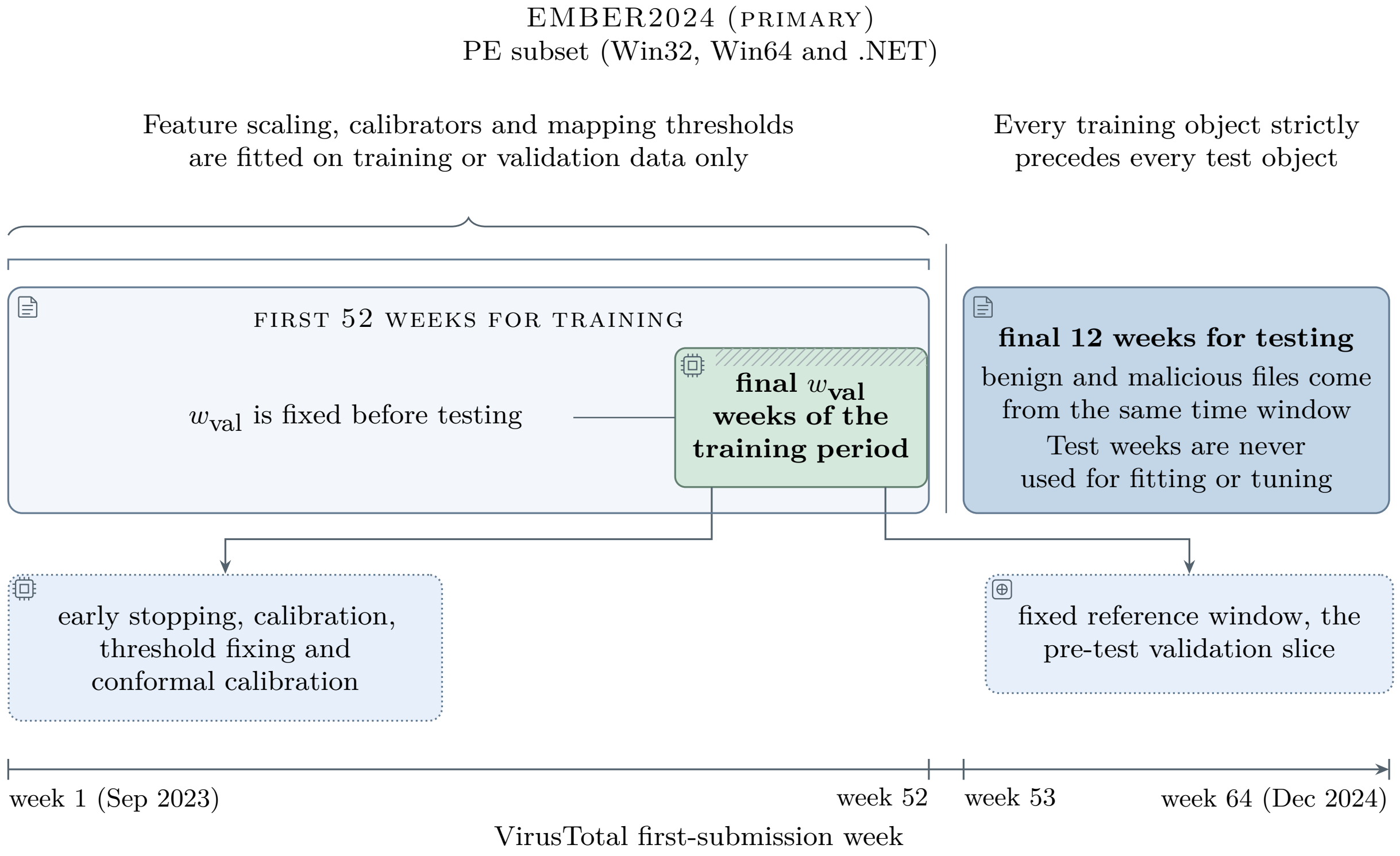

Figure 3: Temporal protocol for the EMBER2024 PE subset. The first 52 weeks form the training period, including its final $w_{\text{val}}$ validation weeks, followed by the final 12 weeks for testing. Validation supplies early stopping, calibration, threshold fixing and conformal calibration; the case-level calibration set and fixed period-level reference window are drawn from the pre-test validation slice. Within each test week, benign and malicious files share the same time window. Test weeks are never used for fitting or tuning.

Table 2: Proposed models and baselines.

| ID | Component | Description | Role |
|---|---|---|---|
| P-GBM | Proposed | LightGBM [58], calibrated | SRQ1 candidate |
| P-MLP | Proposed | Multilayer perceptron (MLP), DL, calibrated | SRQ1 candidate; IG target |
| B0 | End to end | LightGBM in the EMBER2024 benchmark style [33, pp. 5520–5521], retrained on the same split; in our protocol this baseline is uncalibrated and score-only (no explanation, semantic check or drift flag) | End-to-end reference |
| B1 | Prioritization | Logistic regression | SRQ1 family (Holm) |
| B2 | Prioritization | The other model family (MLP vs. LightGBM) | SRQ1 family (Holm) |
| B3 | Prioritization | Random ranking at test prevalence | Floor; descriptive |
| E0 | Explanation | Global LightGBM gain importance applied to every sample | Descriptive |
| E1 | Explanation | Random attribution | Descriptive |
| S0 | Semantic | Permuted feature-to-technique mapping (null) | Descriptive |
| D0 | Drift | No retraining | Descriptive |
| D1 | Drift | Fixed-schedule retraining every 4 test weeks | Descriptive |
| D2 | Drift | Performance-triggered retraining (requires labels) | SRQ4 comparator |

calibration error (ECE) [59] (pitfall P7 in [30, p. 3975]).

- *Prioritization effectiveness (SRQ1).* Workload, the percentage of the weekly queue reviewed to reach 90% malware recall; and precision@$K$ and recall@$K$ for a weekly capacity $K = c \times 45{,}000$ ($c$ is fixed before testing and reported with the results).
- *Explanation quality (SRQ2).* Deletion and insertion AUC for fidelity [60]; top-$k$ Jaccard similarity and rank correlation across seeds and small non-adversarial perturbations for stability; sparsity; runtime per sample; and the degenerate-attribution rate [12, pp. 162–164].
- *Semantic consistency (SRQ3).* Mapping coverage; per-sample precision and recall of the mapped versus the reference (capa `ttps/mbc`) technique set; the per-technique macro average over techniques with $n \geq 30$, with techniques below the threshold listed with their $n$; $\kappa$; the gap to S0; ClarAVy agreement reported per tag, separately and never in the per-technique average; and the AUROC of the per-sample consistency score.
- *Robustness over time (SRQ4).* Per-week PR-AUC; the area under time (AUT) of PR-AUC over 12 weeks [29, pp. 735–736]; detection delay in weeks; false-alarm rate per week; case-level flag precision, the share of conformal-rejected files that the detector misclassified; and PR-AUC recovered after retraining.
- *Triage output (SRQ5).* Requirement coverage of the triage output against a checklist; and worked case walkthroughs.

For SRQ3(b), the population is the high-priority files, defined as the top $K$ of each week. A file in the weekly top $K$ counts as correct when its decision at the validation-fixed threshold matches its label; the share of top-$K$ files below the threshold is reported.

### 5.5 Statistical analysis

- *SRQ1 H0a.* A paired test on ΔPR-AUC (best proposed model minus the stronger of B1 and B2) uses the same test files. The best proposed model is selected on validation data before the test set is first scored. The primary resampling is a moving-block bootstrap over test weeks [61]; a family-cluster bootstrap, which resamples malware files by family and benign files by week, is the sensitivity check. Each replicate recomputes the seed-averaged $\Delta$, with $B$ replicates and 95% percentile confidence intervals (CIs); the block length and $B$ are fixed before testing and reported with the results. Platt scaling leaves ranking metrics unchanged; isotonic regression can alter them through ties.
- *SRQ1 H0b.* McNemar's test [62] at the validation-fixed threshold, in its mid-p form as recommended by Fagerland et al. [63]. A second mid-p McNemar test on benign files only tests false-positive discordance, and FPR is computed on benign files only.
- *Multiplicity.* Holm correction is applied across the model comparisons within each SRQ family [64]. B0, B3, E0, E1, S0, D0 and D1 are descriptive and enter no family.
- *SRQ2 H0a.* A Wilcoxon signed-rank test [65] on paired per-sample fidelity across methods, with Holm correction [64].
- *SRQ2 H0b.* The first and final test weeks are compared with an unpaired Mann–Whitney U test [66] and a bootstrap CI on the difference, because the weeks contain different files.
- *SRQ3(b).* AUROC of the per-sample consistency score with a bootstrap 95% CI; H0 is rejected if the lower bound exceeds 0.5 (one-sided).
- *SRQ4.* Onset is the first test week in which PR-AUC falls below the first-test-week value by more than $\delta$ ($\delta$ is fixed before testing and reported with the results). Delay is compared at a matched false-alarm rate with a Wilcoxon test across seeds and periods, with bootstrap CIs. If no onset occurs, the result is reported as null, with delay undefined and the false-alarm rate still reported; the BODMAS repeat then runs as a separate secondary analysis and is not pooled.
- *SRQ5.* Descriptive. An expert review (Likert ratings; Wilcoxon signed-rank test against 3) is run only if ethics approval is granted; otherwise an analytical evaluation without participants is used.

### 5.6 Ablations

Table 3 lists the ablations.

Table 3: Ablations.

| ID | Change | Question answered |
|---|---|---|
| A1 | Calibration removed (P-GBM vs. B0) | Effect of calibration on ECE, workload and precision@$K$ |
| A2 | Temporal vs. random split | Inflation from temporal snooping |
| A3 | Released ratio vs. subsampled ratio $\pi$ | Base-rate sensitivity of precision@$K$ and workload |
| A4 | Explanation method swap (TreeExplainer, LIME, IG; E0, E1) | Dependence of SRQ3 consistency on the explainer |
| A5 | Top-$k$ values for the mapped set, fixed before testing | Sensitivity of consistency and AUROC to $k$ |
| A6 | Real vs. permuted mapping (S0) | Consistency above chance |
| A7 | Drift signal: conformal only, distribution test only, both | Contribution of each detector |
| A8 | Retraining policy D0, D1, D2, alarm-triggered | Recovery vs. retraining cost |
| A9 | Moving-block vs. family-cluster bootstrap | Robustness of the H0a conclusion |

# 6 Discussion

**Priority versus score.** A detector can rank files well overall yet assign probabilities that do not match how often its high-ranked files are malicious. Calibration decides whether an investigator can take a priority at face value, so the protocol measures ranking quality and calibration separately, on test data that follow the training data in time.

**What the semantic assessment can and cannot show.** The per-sample consistency score measures agreement between mapped attributions and reference tags that are themselves derived from tools and antivirus labels, not from observed behavior. A high score therefore indicates that an explanation is plausible in behavioral terms; it does not show that the decision is correct, and a low score may reflect gaps in the mapping or in the reference tags rather than a wrong explanation. Every agreement figure is reported with its coverage, because files whose top features admit no behavioral reading receive no score.

**Drift signals for reassessment.** Case-level rejection and period-level tests answer different questions: which files the current model should not be trusted on, and when the model as a whole should be reassessed. The EMBER2024 authors suggest that the feature set is resistant to small temporal changes [33, p. 5521], so a weak drift onset is a plausible outcome, which the null-result handling specified before testing and the BODMAS secondary analysis address. Rule-based drift analyses [21, 22] and continual learning such as MADAR [40] are complementary to the flags studied here.

# 7 Limitations and Ethics

## 7.1 Threats to validity

**Construct validity.** Labels come from antivirus consensus, and the consensus threshold is a benchmark-design variable [56]. Labels and reference tags are imperfect (pitfall P2 in [30, p. 3973]);

ClarAVy tags and capa fields are treated as noisy references, not ground-truth behavior. The quality of the capa fields is not assessed. Static features capture only part of a file's behavior [36], and some features have no human meaning. Attributions can reflect spurious artifacts (pitfall P4 in [30, p. 3974]).

**Internal validity.** Tuning leakage is controlled by validation-only fitting and parameters fixed before testing (pitfalls P3 and P5 in [30]). Calibration and conformal sets come from the end of the training period, which may itself be drifting.

**External validity.** The study covers Windows PE files only. Public crawl-based corpora can be sampling-biased (pitfall P1 in [30, p. 3972]); VirusTotal first submissions are not treated here as a random sample of files met in practice. The primary data are a single 64-week collection with a balanced released test set (pitfall P8 in [30]), and public datasets may not match a forensic caseload (pitfall P9 in [30, p. 3976]). The EMBER2024 authors observe that their LightGBM baselines resist multiple weeks of drift [33, p. 5522], so drift onset may be weak.

**Conclusion validity.** Multiple comparisons are handled with Holm correction; results depend on the choice of onset definition; any expert review would have a small $n$; and seeds are averaged with their spread reported.

**Robustness to manipulation.** Explanations and detectors can be manipulated by an adaptive adversary [13, 41, 42] (pitfall P10 in [30, p. 3976]). This work does not evaluate that robustness (Section 3.1). AL-Essa et al. also caution that KernelExplainer sampling variance may contribute to the measured SHAP instability in their comparison with IG [24, p. 23].

### 7.2 Ethics

The study is defensive. It uses only public, pre-extracted static feature vectors and released labels; no malware binaries, raw bytes or disassembly are obtained or distributed, no file is executed, and no third-party system is accessed. The paper and any accompanying material contain no evasion guidance, and feature attributions are reported as aids for investigators. The datasets contain no personal data about human participants. If no expert review is run, SRQ5 is evaluated analytically without participants.

## 8 Conclusion

We presented a framework for Windows PE malware forensic triage that combines a calibrated priority, per-sample explanations, a semantic assessment of those explanations against documented MBC and ATT&CK behaviors, and drift flags at case and period level, all from pre-extracted static features and without executing malware. The evaluation compares each component with simple baselines under a time-ordered protocol specified before testing on EMBER2024, with BODMAS as a secondary dataset.

## References

[1] M. K. Rogers, J. Goldman, R. Mislan, T. Wedge, and S. Debrota, "Computer Forensics Field Triage Process Model," *Journal of Digital Forensics, Security and Law*, vol. 1, no. 2, pp. 19–38, 2006, Art. no. 2, doi: 10.15394/jdfsl.2006.1004.

[2] D. Ucci, L. Aniello, and R. Baldoni, "Survey of machine learning techniques for malware analysis," *Computers & Security*, vol. 81, pp. 123–147, Mar. 2019, doi: 10.1016/j.cose.2018.11.001.

[3] D. Gibert, C. Mateu, and J. Planes, "The rise of machine learning for detection and classification of malware: Research developments, trends and challenges," *Journal of Network and Computer Applications*, vol. 153, Mar. 2020, Art. no. 102526, doi: 10.1016/j.jnca.2019.102526.

[4] H. S. Anderson and P. Roth, "EMBER: An Open Dataset for Training Static PE Malware Machine Learning Models," 2018, arXiv:1804.04637v2, doi: 10.48550/ARXIV.1804.04637.

[5] R. Guidotti, A. Monreale, S. Ruggieri, F. Turini, F. Giannotti, and D. Pedreschi, "A Survey of Methods for Explaining Black Box Models," *ACM Computing Surveys*, vol. 51, no. 5, pp. 1–42, Aug. 2018, Art. no. 93, doi: 10.1145/3236009.

[6] P. Shingade, P. Dhotre, and P. Kharat, "Enhancing Trust and Transparency in Digital Forensics and Incident Response using Explainable AI (XAI): An Extensive Survey," in *2025 IEEE International Conference on Blockchain and Distributed Systems Security (ICBDS)*, IEEE, Oct. 2025, pp. 1–5, doi: 10.1109/ICBDS67396.2025.11377825.

[7] S. W. Hall, A. Sakzad, and K.-K. R. Choo, "Explainable artificial intelligence for digital forensics," *WIREs Forensic Science*, vol. 4, no. 2, Mar. 2022, Art. no. e1434, doi: 10.1002/wfs2.1434.

[8] M. T. Ribeiro, S. Singh, and C. Guestrin, "'Why Should I Trust You?': Explaining the Predictions of Any Classifier," in *Proceedings of the 22nd ACM SIGKDD International Conference on Knowledge Discovery and Data Mining*, ser. KDD '16, ACM, Aug. 2016, pp. 1135–1144, doi: 10.1145/2939672.2939778.

[9] S. M. Lundberg and S.-I. Lee, "A Unified Approach to Interpreting Model Predictions," in *Advances in Neural Information Processing Systems*, vol. 30, 2017. [Online]. Available: https://proceedings.neurips.cc/paper/2017/hash/8a20a8621978632d76c43dfd28b67767-Abstract.html

[10] H. Manthena, S. Shajarian, J. C. Kimmell, M. Abdelsalam, S. Khorsandroo, and M. Gupta, "Explainable Artificial Intelligence (XAI) for Malware Analysis: A Survey of Techniques, Applications, and Open Challenges," *IEEE Access*, vol. 13, pp. 61 611–61 640, 2025, doi: 10.1109/ACCESS.2025.3555926.

[11] M. Sundararajan, A. Taly, and Q. Yan, "Axiomatic Attribution for Deep Networks," in *Proceedings of the 34th International Conference on Machine Learning*, ser. Proceedings of Machine Learning Research, vol. 70, PMLR, 2017, pp. 3319–3328. [Online]. Available: https://proceedings.mlr.press/v70/sundararajan17a.html

[12] A. Warnecke, D. Arp, C. Wressnegger, and K. Rieck, "Evaluating Explanation Methods for Deep Learning in Security," in *2020 IEEE European Symposium on Security and Privacy (EuroS&P)*, IEEE, Sep. 2020, pp. 158–174, doi: 10.1109/eurosp48549.2020.00018.

[13] M. Noppel and C. Wressnegger, "SoK: Explainable Machine Learning in Adversarial Environments," in *2024 IEEE Symposium on Security and Privacy (SP)*, IEEE, May 2024, pp. 2441–2459, doi: 10.1109/sp54263.2024.00021.

[14] Microsoft, “PE Format,” Microsoft Learn, Win32 apps documentation, Sep. 2026, Page dated 10 September 2026. Accessed: Nov. 11, 2026. [Online]. Available: https://learn.microsoft.com/en-us/windows/win32/debug/pe-format

[15] B. E. Strom, A. Applebaum, D. P. Miller, K. C. Nickels, A. G. Pennington, and C. B. Thomas, “MITRE ATT&CK: Design and Philosophy,” The MITRE Corporation, Tech. Rep., 2020, Originally published July 2018; revised March 2020. [Online]. Available: https://www.mitre.org/news-insights/publication/mitre-attck-design-and-philosophy

[16] MBC Project, “Malware Behavior Catalog (MBC),” GitHub repository, version 3.2 (release v3.2, Dec. 9, 2024), 2024, Repository tag v3.2, commit 0aa720aa318924b8b102107ee0f004b5993bc8ad. Documentation: README.md (purpose and ATT&CK links); yfaq/README.md (vocabulary and mapping guidance); defense-evasion/obfuscated-files-or-information.md (E1027 related techniques). README/FAQ retain internal v3.1 labels. Accessed: Nov. 9, 2026. [Online]. Available: https://github.com/MBCProject/mbc-markdown/tree/v3.2

[17] P. Švec, Š. Balogh, M. Homola, J. Kľuka, T. Bisták, and P. Anthony, “Semantic data representation for explainable Windows malware detection models,” *Journal of Computer Virology and Hacking Techniques*, vol. 22, no. 1, Jun. 2026, Art. no. 54, doi: 10.1007/s11416-026-00635-y.

[18] E. Downing, Y. Mirsky, K. Park, and W. Lee, “DeepReflect: Discovering Malicious Functionality through Binary Reconstruction,” in *30th USENIX Security Symposium (USENIX Security 21)*, 2021, pp. 3469–3486. [Online]. Available: https://www.usenix.org/conference/usenixsecurity21/presentation/downing

[19] Y. He, J. Lei, Z. Qin, K. Ren, and C. Chen, “Combating Concept Drift with Explanatory Detection and Adaptation for Android Malware Classification,” in *Proceedings of the 2025 ACM SIGSAC Conference on Computer and Communications Security*, ser. CCS ’25, ACM, Nov. 2025, pp. 978–992, doi: 10.1145/3719027.3744792.

[20] P. Anthony, K. R. Galadima, M. Homola, Š. Balogh, J. Kľuka, P. Švec, R. Ploszek, and Z. Adams, “PaTEx: Decision-Aware Partitioning for Tree-Structured Explanations in Neural Malware Classification,” *Procedia Computer Science*, vol. 292, pp. 5879–5890, 2026, 30th International Conference on Knowledge-Based and Intelligent Information & Engineering Systems (KES 2026), doi: 10.1016/j.procs.2026.10.159.

[21] T. Kalný, M. Jureček, and M. Stamp, “Detecting Concept Drift in Evolving Malware Families Using Rule-Based Classifier Representations,” in *Proceedings of the 23rd International Conference on Security and Cryptography (SECRYPT 2026), Volume 1*, SCITEPRESS, 2026, pp. 922–931, doi: 10.5220/0015184000004103.

[22] O. Jurečková and M. Jureček, “Detecting and Explaining Malware Family Evolution Using Rule-Based Drift Analysis,” in *Proceedings of the 12th International Conference on Information Systems Security and Privacy (ICISSP 2026), Volume 1*, SCITEPRESS, 2026, pp. 366–374, doi: 10.5220/0014638000004061.

[23] N. H. Guldemir, O. Olukoya, and J. Martínez-del-Rincón, “Explaining concept drift in Windows PE malware with integrated gradients and behavioral analysis,” *Cybersecurity*, vol. 9, no. 1, 2026, Art. no. 254, doi: 10.1186/s42400-026-00691-3.

[24] M. AL-Essa, M. Atoum, M. Imran, W. A. ALi, N. Alassaf, and M. S. Alsaadi, “Evaluating the robustness of explainable AI methods to adversarial attacks in intrusion and malware detection,” *Discover Artificial Intelligence*, vol. 6, no. 1, 2026, Art. no. 1811, doi: 10.1007/s44163-026-02587-0.

[25] The MITRE Corporation, "MAEC 5.0 (Malware Attribute Enumeration and Characterization) Release," MAEC Project Documentation, 2017, Accessed: Oct. 21, 2026. [Online]. Available: `https://maecproject.github.io/releases/5.0/`

[26] J. Gama, I. Žliobaitė, A. Bifet, M. Pechenizkiy, and A. Bouchachia, "A survey on concept drift adaptation," *ACM Computing Surveys*, vol. 46, no. 4, pp. 1–37, Mar. 2014, doi: `10.1145/2523813`.

[27] A. Ponte, D. Gibert, M. Kozak, D. Trizna, M. Pintor, B. Biggio, F. Roli, and L. Demetrio, "EXE-Bench: Ranking the Tradeoffs of AI-Based Windows Malware Detectors for Real-World Usability," *IEEE Transactions on Information Forensics and Security*, vol. 21, pp. 9007–9020, 2026, doi: `10.1109/tifs.2026.3734783`.

[28] Y. Chen, Z. Ding, and D. Wagner, "Continuous Learning for Android Malware Detection," in *32nd USENIX Security Symposium*, 2023, pp. 1127–1144. [Online]. Available: `https://www.usenix.org/conference/usenixsecurity23/presentation/chen-yizheng`

[29] F. Pendlebury, F. Pierazzi, R. Jordaney, J. Kinder, and L. Cavallaro, "TESSERACT: Eliminating Experimental Bias in Malware Classification across Space and Time," in *28th USENIX Security Symposium*, 2019, pp. 729–746. [Online]. Available: `https://www.usenix.org/conference/usenixsecurity19/presentation/pendlebury`

[30] D. Arp, E. Quiring, F. Pendlebury, A. Warnecke, F. Pierazzi, C. Wressnegger, L. Cavallaro, and K. Rieck, "Dos and Don'ts of Machine Learning in Computer Security," in *31st USENIX Security Symposium*, 2022, pp. 3971–3988. [Online]. Available: `https://www.usenix.org/conference/usenixsecurity22/presentation/arp`

[31] F. Barbero, F. Pendlebury, F. Pierazzi, and L. Cavallaro, "Transcending TRANSCEND: Revisiting Malware Classification in the Presence of Concept Drift," in *2022 IEEE Symposium on Security and Privacy (SP)*, IEEE, May 2022, pp. 805–823, doi: `10.1109/sp46214.2022.9833659`.

[32] L. Yang, W. Guo, Q. Hao, A. Ciptadi, A. Ahmadzadeh, X. Xing, and G. Wang, "CADE: Detecting and Explaining Concept Drift Samples for Security Applications," in *30th USENIX Security Symposium*, 2021, pp. 2327–2344. [Online]. Available: `https://www.usenix.org/conference/usenixsecurity21/presentation/yang-limin`

[33] R. J. Joyce, G. Miller, P. Roth, R. Zak, E. Zaresky-Williams, H. Anderson, E. Raff, and J. Holt, "EMBER2024 - A Benchmark Dataset for Holistic Evaluation of Malware Classifiers," in *Proceedings of the 31st ACM SIGKDD Conference on Knowledge Discovery and Data Mining V.2*, ser. KDD '25, ACM, Aug. 2025, pp. 5516–5526, doi: `10.1145/3711896.3737431`.

[34] L. Yang, A. Ciptadi, I. Laziuk, A. Ahmadzadeh, and G. Wang, "BODMAS: An Open Dataset for Learning based Temporal Analysis of PE Malware," in *2021 IEEE Security and Privacy Workshops (SPW)*, IEEE, May 2021, pp. 78–84, doi: `10.1109/spw53761.2021.00020`.

[35] D. Gibert, N. Totosis, C. Patsakis, Q. Le, and G. Zizzo, "Assessing the impact of packing on static machine learning-based malware detection and classification systems," *Computers & Security*, vol. 156, 2025, Art. no. 104495, doi: `10.1016/j.cose.2025.104495`.

[36] A. Moser, C. Kruegel, and E. Kirda, "Limits of Static Analysis for Malware Detection," in *Twenty-Third Annual Computer Security Applications Conference (ACSAC 2007)*, IEEE, Dec. 2007, pp. 421–430, doi: `10.1109/ACSAC.2007.21`.

[37] E. Raff, J. Barker, J. Sylvester, R. Brandon, B. Catanzaro, and C. Nicholas, "Malware Detection by Eating a Whole EXE," in *The Workshops of the Thirty-Second AAAI Conference on Artificial Intelligence*, 2018, pp. 268–276. [Online]. Available: https://arxiv.org/abs/1710.09435

[38] R. Jordaney, K. Sharad, S. K. Dash, Z. Wang, D. Papini, I. Nouretdinov, and L. Cavallaro, "Transcend: Detecting Concept Drift in Malware Classification Models," in *26th USENIX Security Symposium*, 2017, pp. 625–642. [Online]. Available: https://www.usenix.org/conference/usenixsecurity17/technical-sessions/presentation/jordaney

[39] A. S. Li, A. Iyengar, A. Kundu, and E. Bertino, "Revisiting Concept Drift in Windows Malware Detection: Adaptation to Real Drifted Malware with Minimal Samples," in *Proceedings 2025 Network and Distributed System Security Symposium*, ser. NDSS 2025, Internet Society, 2025, doi: 10.14722/ndss.2025.240830.

[40] M. S. Rahman, S. Coull, Q. Yu, and M. Wright, "MADAR: Efficient Continual Learning for Malware Analysis with Distribution-Aware Replay," in *Proceedings of the 2025 Conference on Applied Machine Learning for Information Security (CAMLIS)*, ser. Proceedings of Machine Learning Research, vol. 299, PMLR, 2025, pp. 265–291. [Online]. Available: https://proceedings.mlr.press/v299/rahman25a.html

[41] X. Zhang, N. Wang, H. Shen, S. Ji, X. Luo, and T. Wang, "Interpretable Deep Learning under Fire," in *29th USENIX Security Symposium*, 2020, pp. 1659–1676. [Online]. Available: https://www.usenix.org/conference/usenixsecurity20/presentation/zhang-xinyang

[42] G. Severi, J. Meyer, S. Coull, and A. Oprea, "Explanation-Guided Backdoor Poisoning Attacks Against Malware Classifiers," in *30th USENIX Security Symposium*, 2021, pp. 1487–1504. [Online]. Available: https://www.usenix.org/conference/usenixsecurity21/presentation/severi

[43] F. Alotaibi, E. Goodbrand, and S. Maffeis, "Deep Learning from Imperfectly Labeled Malware Data," in *Proceedings of the 2025 ACM SIGSAC Conference on Computer and Communications Security*, ACM, 2025, pp. 3990–4004, doi: 10.1145/3719027.3765197.

[44] J. Mink, H. Benkraouda, L. Yang, A. Ciptadi, A. Ahmadzadeh, D. Votipka, and G. Wang, "Everybody's Got ML, Tell Me What Else You Have: Practitioners' Perception of ML-Based Security Tools and Explanations," in *2023 IEEE Symposium on Security and Privacy (SP)*, IEEE, 2023, pp. 2068–2085, doi: 10.1109/sp46215.2023.10179321.

[45] J. C. Platt, "Probabilities for SV Machines," in *Advances in Large-Margin Classifiers*, A. J. Smola, P. Bartlett, B. Schölkopf, and D. Schuurmans, Eds., The MIT Press, 2000, pp. 61–74, doi: 10.7551/mitpress/1113.003.0008.

[46] B. Zadrozny and C. Elkan, "Transforming Classifier Scores into Accurate Multiclass Probability Estimates," in *Proceedings of the Eighth ACM SIGKDD International Conference on Knowledge Discovery and Data Mining (KDD '02)*, ACM, Jul. 2002, pp. 694–699, doi: 10.1145/775047.775151.

[47] S. M. Lundberg, G. Erion, H. Chen, A. DeGrave, J. M. Prutkin, B. Nair, R. Katz, J. Himmelfarb, N. Bansal, and S.-I. Lee, "From Local Explanations to Global Understanding with Explainable AI for Trees," *Nature Machine Intelligence*, vol. 2, no. 1, pp. 56–67, Jan. 2020, doi: 10.1038/s42256-019-0138-9.

[48] J. Cohen, "A Coefficient of Agreement for Nominal Scales," *Educational and Psychological Measurement*, vol. 20, no. 1, pp. 37–46, Apr. 1960, doi: 10.1177/001316446002000104.

[49] Mandiant FLARE, "capa: The FLARE team's open-source tool to identify capabilities in executable files," GitHub repository, Accessed: Nov. 4, 2026. [Online]. Available: https://github.com/mandiant/capa

[50] R. Yumlembam, B. Issac, and S. M. Jacob, "Enhancing decision-making in windows PE malware classification during dataset shifts with uncertainty estimation," *Knowledge-Based Systems*, vol. 331, 2026, Art. no. 114723, doi: 10.1016/j.knosys.2025.114723.

[51] N. Smirnov, "Table for Estimating the Goodness of Fit of Empirical Distributions," *The Annals of Mathematical Statistics*, vol. 19, no. 2, pp. 279–281, Jun. 1948, doi: 10.1214/aoms/1177730256.

[52] B. Yurdakul and J. Naranjo, "Statistical Properties of the Population Stability Index," *The Journal of Risk Model Validation*, vol. 14, no. 4, pp. 89–100, Dec. 2020, doi: 10.21314/JRMV.2020.227.

[53] A. Gretton, K. M. Borgwardt, M. J. Rasch, B. Schölkopf, and A. Smola, "A Kernel Two-Sample Test," *Journal of Machine Learning Research*, vol. 13, pp. 723–773, 2012. [Online]. Available: https://jmlr.org/papers/v13/gretton12a.html

[54] FutureComputing4AI, "EMBER2024," GitHub repository README, commit 0ef753e8, 2025. Accessed: Oct. 28, 2026. [Online]. Available: https://github.com/FutureComputing4AI/EMBER2024/tree/0ef753e81d98bf209f71b03cd331dfc190b5b54d

[55] R. J. Joyce, D. Everett, M. Fuchs, E. Raff, and J. Holt, "ClarAVy: A Tool for Scalable and Accurate Malware Family Labeling," in *Companion Proceedings of the ACM on Web Conference 2025*, ser. WWW '25, ACM, May 2025, pp. 277–286, doi: 10.1145/3701716.3715212.

[56] T.-T. Huynh, V.-Q. Trinh, D.-T. Huynh, and N.-H. Le, "Auditing VirusTotal Consensus Thresholds in Malware Benchmarks: A Reproducible Security-Measurement Study on EMBER2024," *Journal of Information Security and Applications*, vol. 103, Dec. 2026, Art. no. 104659, doi: 10.1016/j.jisa.2026.104659.

[57] Y. He, J. Lou, Z. Qin, and K. Ren, "FINER: Enhancing State-of-the-art Classifiers with Feature Attribution to Facilitate Security Analysis," in *Proceedings of the 2023 ACM SIGSAC Conference on Computer and Communications Security*, ser. CCS '23, ACM, Nov. 2023, pp. 416–430, doi: 10.1145/3576915.3616599.

[58] G. Ke, Q. Meng, T. Finley, T. Wang, W. Chen, W. Ma, Q. Ye, and T.-Y. Liu, "LightGBM: A Highly Efficient Gradient Boosting Decision Tree," in *Advances in Neural Information Processing Systems 30 (NeurIPS 2017)*, Curran Associates, Inc., 2017. [Online]. Available: https://proceedings.neurips.cc/paper_files/paper/2017/hash/6449f44a102fde848669bdd9eb6b76fa-Abstract.html

[59] C. Guo, G. Pleiss, Y. Sun, and K. Q. Weinberger, "On Calibration of Modern Neural Networks," in *Proceedings of the 34th International Conference on Machine Learning (ICML)*, ser. Proceedings of Machine Learning Research, vol. 70, 2017, pp. 1321–1330. [Online]. Available: https://proceedings.mlr.press/v70/guo17a.html

[60] V. Petsiuk, A. Das, and K. Saenko, "RISE: Randomized Input Sampling for Explanation of Black-box Models," in *British Machine Vision Conference (BMVC)*, 2018. [Online]. Available: https://bmva-archive.org.uk/bmvc/2018/contents/papers/1064.pdf

[61] H. R. Künsch, "The Jackknife and the Bootstrap for General Stationary Observations," *The Annals of Statistics*, vol. 17, no. 3, pp. 1217–1241, Sep. 1989, doi: 10.1214/aos/1176347265.

[62] Q. McNemar, "Note on the Sampling Error of the Difference between Correlated Proportions or Percentages," *Psychometrika*, vol. 12, no. 2, pp. 153–157, Jun. 1947, doi: 10.1007/BF02295996.

[63] M. W. Fagerland, S. Lydersen, and P. Laake, "The McNemar Test for Binary Matched-Pairs Data: Mid-p and Asymptotic Are Better than Exact Conditional," *BMC Medical Research Methodology*, vol. 13, no. 1, Jul. 2013, Art. no. 91, doi: 10.1186/1471-2288-13-91.

[64] S. Holm, "A Simple Sequentially Rejective Multiple Test Procedure," *Scandinavian Journal of Statistics*, vol. 6, no. 2, pp. 65–70, 1979. [Online]. Available: https://www.jstor.org/stable/4615733

[65] F. Wilcoxon, "Individual Comparisons by Ranking Methods," *Biometrics Bulletin*, vol. 1, no. 6, pp. 80–83, Dec. 1945, doi: 10.2307/3001968.

[66] H. B. Mann and D. R. Whitney, "On a Test of Whether One of Two Random Variables Is Stochastically Larger than the Other," *The Annals of Mathematical Statistics*, vol. 18, no. 1, pp. 50–60, Mar. 1947, doi: 10.1214/aoms/1177730491.